\documentclass[11pt, oneside]{article}   	
\usepackage{geometry}                		
\usepackage{graphicx}				
\usepackage{amssymb}

\usepackage{amsmath}
\usepackage{graphicx}				
\usepackage{amssymb}

\usepackage{hyperref}
\usepackage{authblk}

\newcommand{\gst}{Ge$_2$Sb$_2$Te$_5$~}
\newcommand{\SbS}{Sb$_2$S$_3$~}
\newcommand{\SbSns}{Sb$_2$S$_3$}
\newcommand{\sbte}{Sb$_2$Te$_3$~}
\newcommand{\sbtens}{Sb$_2$Te$_3$}
\newcommand{\gstns}{Ge$_2$Sb$_2$Te$_5$}
\newcommand{\sn}{Si$_3$N$_4$~}

\newcommand{\zsso}{($ZnS$)$_{0.8}$)-($Si$O$_{2}$)$_{0.2}$~}

\newcommand{\celsiusns}{$^\circ$C}

\usepackage{subfigure}
\usepackage{color, soul}
\usepackage{textcomp} 
\usepackage{graphicx}

\title{Phase Change Material Photonics}
\author[1]{Robet E Simpson}
\author[2]{Tun Cao}
\affil[1]{Singapore University of Technology, 8 Somapah Road,  487372, Singapore}
\affil[2]{Dalian University of Technology, Dalian 116024, China}

\date{}							

\begin{document}
\maketitle

\section{Introduction}
In the last decade phase change materials (PCM) research has switched from practical application in optical data storage toward electrical phase change random access memory technologies (PCRAM).
As these devices are commercialised, we expect the research direction to switch once again toward electrical-photonic devices.

 When phase change material devices are heated using nanosecond pulses, the material can be SET into the crystal state, and RESET into the amorphous state, which changes the electrical resistivity by orders of magnitude and radically changes the PCM optical constants. 
It is this optical change that is exploited in phase change photonics.
The objective of this chapter is to introduce the concepts in PCM-tuned photonics.
We will start by reviewing the key works in the field, before concentrating on Metal-Dielectric-Metal (MDM) structures.
We will discuss how to design tuneable-MDM photonics devices, their advantages, and their limitations. 
Finally we will discuss new materials for phase change photonics.  

\subsection{Motivation}

The properties of traditional optical materials, such as glasses, are usually fixed. 
Therefore, the only way to change the behaviour of optical devices composed from these fixed optical materials is by precisely altering the device geometry.
Since the wavelength of light is sub-micron,   the geometry of these optical structures must be precisely controlled, which makes tuneable optical systems expensive.
However, more modern materials with tuneable optical properties, such as liquid crystals, electrochromics, and phase change  materials  can be used to control the propagation of light in a structure with a fixed geometry.
Fixing the geometry of a tuneable optical system reduces its complexity and should lower the manufacturing cost.
Applications of tuneable photonics include telecommunications, displays, optical computing, and sensing, which employ tuneable monochromators, modulators, and lenses. 
Clearly, reducing the manufacturing cost of these devices will also make these  technologies more affordable to the end users.

Liquid crystals and electrochromic materials tend to exhibit a rather small change in their optical constants when switched.
Moreover, their switching speed tends to be at best tens of microseconds,  and when the switching energy source is removed, these materials revert back to their default optical state. 
I.e. they are `momentary' switches.
In contrast, phase change materials can switch on a sub-nanosecond time scale\cite{Waldecker2015nmat, Loke12Sci}, and their optical state is non-volatile. 
They retain their optical state even when no energy is supplied.
I.e. they are 'latch' switches.
This makes phase change materials particularly attractive for low energy optical devices that need to be switched quickly.

\section{Background}
The application of PCMs to optics and photonics was  limited  to optical data storage until the advent of nanophotonics in the 2000s.
Nanophotonics requires only small quantities of tuneable material to  control the properties of the nanoscale optical structures.
This is important because the energy used to thermally switch PCMs decreases with  volume\cite{pirovano03}.
From a thermal design perspective,  the smaller  the volume of PCM, the easier it is to quench at the rates necessary for re-amorphisation, which is required for reversible switching.
 
 In 2006, Strand et al used simulations to demonstrate the potential of  phase change data storage materials for tuning the properties of nanophotonic devices\cite{Strand}.
Two very different devices were simulated: (1) a switchable photonic crystal add-drop multiplexer, and (2) a reflective phase array.
Both devices were designed to operate at a wavelength of 1550~nm, which is at the center of the telecoms C-band.
The photonic crystal consisted of arrays of holes drilled into the high refractive index PCM.
The optimum design employed a 225~nm phase change layer with arrays of holes that were also 225~nm in diameter. 
The beam steering phase array  required repeating cells which were 500~nm wide squares. 
Each cell was then sub-divided into a 7-by-7 array or PCM pillars. 
Therefore, the largest pillar  had a diameter of 71.5~nm.
These features are, therefore, an order of magnitude smaller than the wavelength of light.
Accurately patterning these periodic structures over a large area is extremely challenging. 
Indeed,  the \gst-based all-dielectrics phase array beam steering devices were only demonstrated in 2019, some 16 years after the simulation paper\cite{Galarreta2019axiv}, whilst the PCM photonic crystal add-drop mulitplexer has not yet been demonstrated. 

\begin{figure}[htbp]
   \centering
   \subfigure[]{\includegraphics[width=0.45\textwidth]{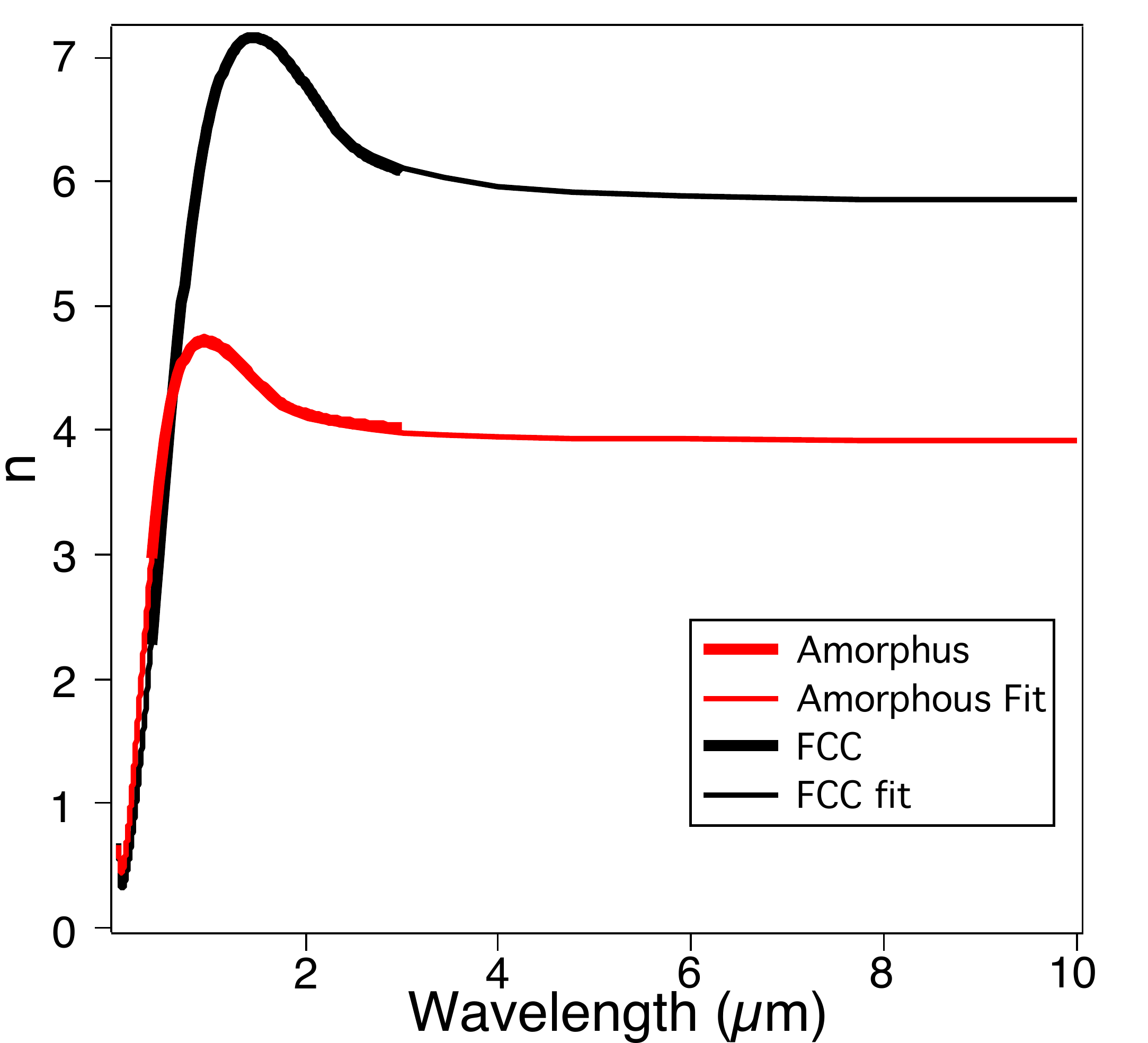} }   
   \subfigure[]{\includegraphics[width=0.45\textwidth]{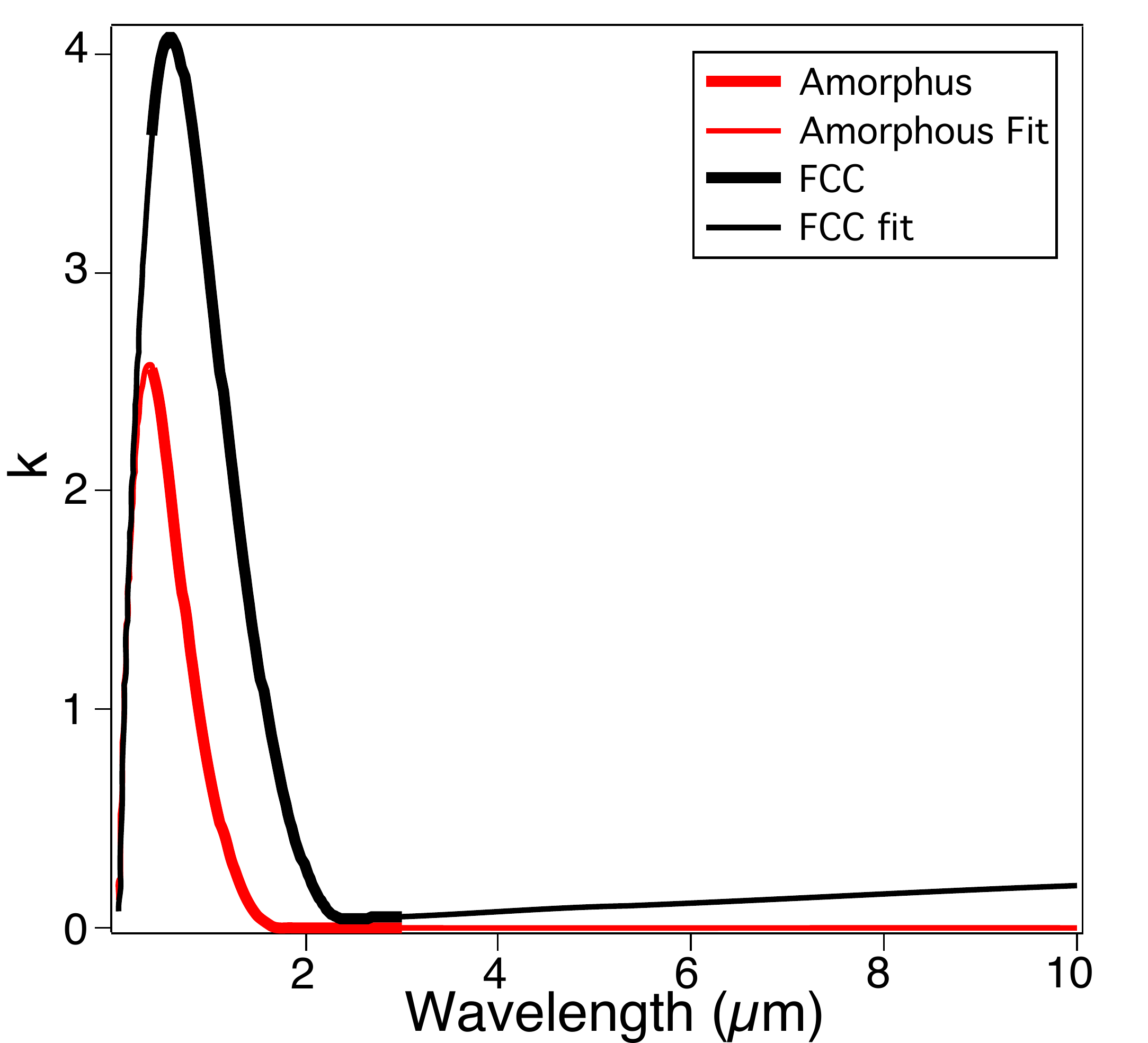} }
   \caption{Complex refractive index of \gst  as a function of wavelength. 
   The real component, Re(n), is shown in (a), whilst the extinction coefficient, k, is  shown in (b). }
   \label{gstnk}
\end{figure}

Strand  et al. also realised that the conventional phase change data storage materials along the \sbtens--GeTe pseudo-binary tie-line have a high optical absorption at  telecommunications wavelengths.
Figure \ref{gstnk} shows the complex refractive index of \gstns.
As can be seen, the extinction coefficient, k, which is related to optical absorption, is large for wavelengths shorter than 2000~nm.
Since Strand's devices need to be non-lossy, the PCM must have an absorption lower than \gstns.
To achieve this they alloyed small amounts of unnamed elements with \gst to lower k  below 0.01 at a wavelength of 1550~nm\cite{Strand}.
The high optical absorption of phase change materials is a key problem that needs to be overcome for PCM photonics to have  real practical applications. 
We will discuss this further in section \ref{sec_materials}.

The current phase change photonics  research landscape  can be split into three themes: (1) On-chip photonics, (2) Metasurfaces, and (3) Materials Engineering. 
It is interesting that these three themes were all addressed in the seminal simulations by Strand et al\cite{Strand}. 

The GeTe-\sbte{} psuedo-binary phase change alloys were developed by the Panasonic (formerly known as Matsushita Electric Industrial Co., Ltd) for data storage applications\cite{yamada1987hso, yamada91}; a Japanese  corporation.
Therefore, it is seems to be no coincidence that the first prototype \gst-tuned on-chip photonics devices emerged from  a Japanese research group.
Ikuma et al. first showed that  an optical waveguide \gst-based gate-switch could attenuate 1550~nm signals by 12.5~dB\cite{Ikuma10EL}.
However, reversible switching was not demonstrated.
Nonetheless this was an important demonstration that  high extinction values are possible at 1550~nm using phase change data storage materials.

\section{On chip tuneable photonics}
\begin{figure}[htbp] 
   \centering
   \includegraphics[width=\textwidth]{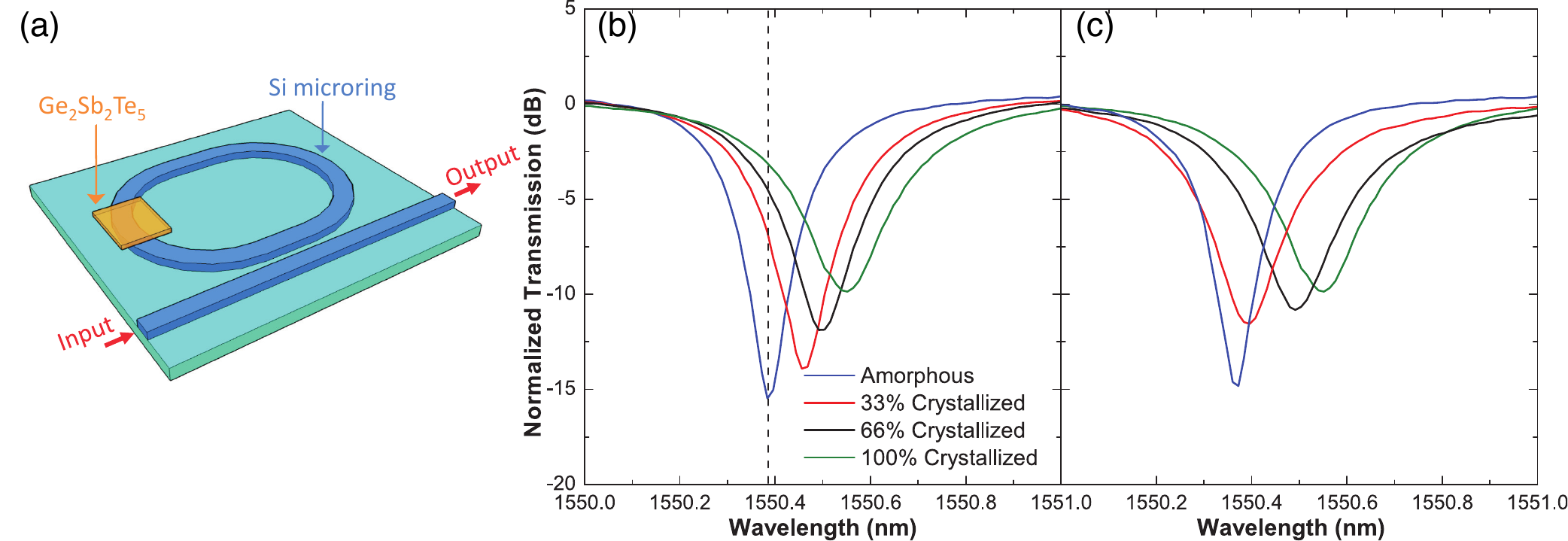} 
   \caption{\gst-tuned ring resonator. (a) A schematic of the resonator showing the \gst patch on the Si ring waveguide. 
   (b) The ring's measured resonant wavelength is shifted from 1550.38~nm to 1550.55~nm when the \gst patch is progressively crystallised.
   (c) The reversible transition back to the amorphous state causes the ring's resonant frequency to switch back to  1550.38~nm.
   Figure adapted from ref. \cite{rude2013APL}, AIP Publishing}
   \label{RudeRR}
\end{figure}

Reversible switching in PCM-tuned on-chip photonics devices was first demonstrated by Rude et al.\cite{rude2013APL} using a Si microring Resonator on-chip photonics device.
Rather than exploiting the change in the  imaginary part ($Im(n)$) of the refractive index, which is related to the optical absorption, this resonator device exploited the change in the real part of the refractive index ($Re(n)$).
The idea was to switch the optical path length of the resonator by increasing the Re{n} of a \gst patch on the Si resonator waveguide, which in turn would change the the resonant wavelength of the device. 
Fig. \ref{RudeRR}(a) shows the schematic of the device.
The resonant wavelength of the prototype devices could be reversibly switched from 1550.38~nm to 1550.55~nm.
Moreover, changing the fraction of  crystallised \gst  allowed the resonance to be tuned to any wavelength between 1550.38~nm to 1550.55~nm, see figure \ref{RudeRR}(b-c) 
The tuning was accomplished using an external 975~nm pump laser to heat the \gst patch. 
The advantage of tuneable resonators is that they can be used to control the phase  and amplitude of the wave emerging from the resonator; thus tuneable interference is possible.
Indeed, this device also showed a 12~dB extinction ratio by exploiting the switch in Re{n} to destructively interfere the  1550.38~nn signal.
A similar \gst-tuned ring resonator that used a \sn waveguide rather than Si was demonstrated in the same year by Rios et al\cite{rios2013AM}. 
Generally, absorption inside resonators  decreases the amount of energy resonating, which concomitantly increases the width of the resonance.
Since \gst is relatively absorbing at 1550~nm, the ring resonators are  relatively lossy and have broad resonances (Low Q-factor).
Thus new low-loss PCMs are required to tune telecommunication wavelength ring resonators.

More recently, Rios et al. demonstrated  that the transmitted intensity of 1550~nm light through a \sn waveguide could be reversibly switched by heating the \gst with a laser pump pulse\cite{Rios15NP}.
Unlike the previous switching demonstrations, the pump pulse was also sent through the waveguide. 
The \gst switches because  a significant fraction of the guided mode is outside of the waveguide, and since the refractive index of \gst is much larger than \sn, the electric field of the propagating mode couples into the \gstns.
A stepwise partial recrystallization scheme was employed because  amorphous \gst  has a much lower refractive index than the crystalline state; hence less energy is evanescently coupled  to the amorphous \gst patch, which makes single-shot crystallisation  challenging.

\begin{figure}[htbp]
   \centering
   \includegraphics[width=0.6\textwidth]{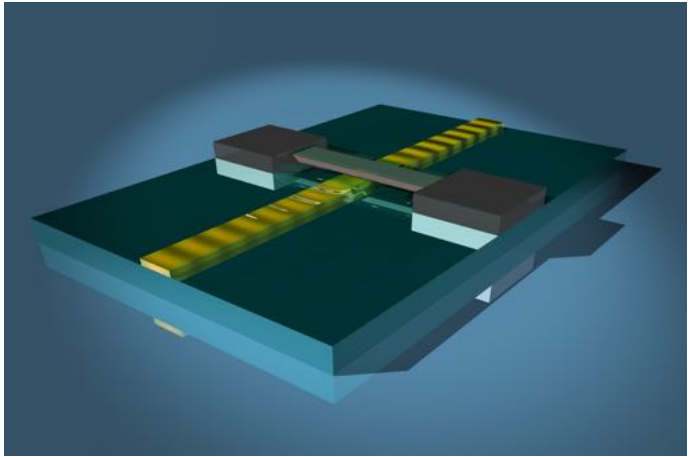} 
   \caption{Schematic of the plasmonic random access memory concept \cite{Rude15ACSP}.}
   \label{plasmoram}
\end{figure}

Aside from switching the transmission of light in dielectric waveguides, surface plasmon polaritons (SPPs) and surface phonon polaritons (SPhPs) can also be switched using \gst\cite{Rude15ACSP, Li16nmat}.
The Drude model states that free electrons oscillate $180^\circ$ out of phase with the driving electric field, and therefore the dielectric permittivity of most metals is negative.
Plasmons are collective oscillations of these free electrons at a well defined resonant frequency.
SPPs are collective oscillations of free electrons at a metal-dielectric interfaces.
In contrast,  SPhPs can be excited by coupling light  to polar crystals, which produces collective lattice vibrations at a well defined frequency. 
Both effects involve the interaction of electromagnetic radiation under the condition that $|\epsilon_1|>|\epsilon_2|$, where  the dielectric function is defined as $\epsilon=\epsilon_1 +i\epsilon_2$.
At visible frequencies SPPs tend to occur when light interacts with metallic nanostructures.
Whilst mid-infrared frequencies are usually used to excite SPhPs.
Both SPPs and SPhPs are highly confined guided modes, but SPhPs tend to have a much longer lifetime and are less lossy.
Rude et al. showed that 1550~nm SPPs   guided along a Au/SiO$_2$ interface can be modulated by changing the structural state of a \gst section, which is patterned on the SPP waveguide\cite{Rude15ACSP}.
Conceivably, this type of device could be used to read the state of nanoscale electrical phase change random access memory devices using light,  which would allow faster communication across a computer network.
A schematic of the structure is shown in figure \ref{plasmoram}.
Li et al. demonstrated that thin \gst films  on a polar crystal, such as quatz, can  support M-IR SPhPs\cite{Li16nmat}.
By laser crystallising  two dimensional structures into the \gst, high Q-factor mid-Infrared SPhP resonators were possible. 
Moreover, by switching the state of the \gst the resonant wavelength could be tuned.

\section{Planar thin film tuneable filters}
\label{planar}
\begin{figure}[htbp]
\begin{center}
\includegraphics[width=\textwidth]{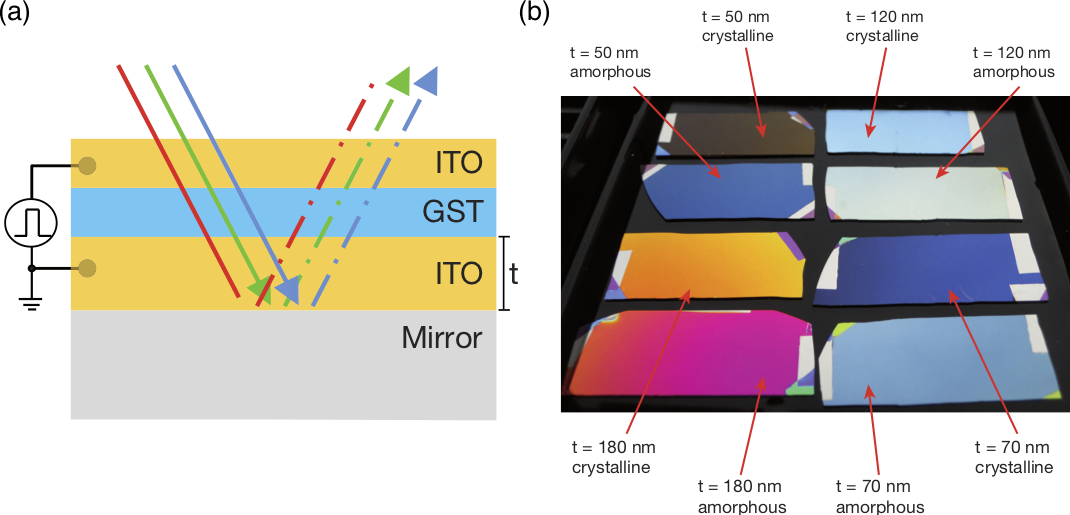}
\caption{(a) Schematic  of the thin-film ITO/\gstns/ITO stack. 
(b) Four different films with similar layered structures  consisting of 10~nm~ITO/7~nm~\gst/$t$~nm~ITO/100~nm~Pt with different thicknesses of $t$. 
After crystallising the \gstns, the colour of the film changes.
Figure adapted from reference \cite{Hosseini2014Nature},  Nature Publishing Group }
\label{hosseini}
\end{center}
\end{figure}

PCMs have been used to tune the reflectivity spectrum of planar filters.  
Hosseini showed that a simple  structure consisting of ITO/\gst/ITO  layers  can produce strong  resonances in the visible spectrum\cite{Hosseini2014Nature}. 
The reflected colour is  defined by the thicknesses of the ITO and \gst layers.
Essentially, this structure forms  a thin film Fabry-Perot etalon.
To ensure  colour  vibrancy, the \gst films must be extremely thin so that energy is only absorbed when the wavelength of the incident light is resonant with the etalon. 
For example, in the Fig. \ref{hosseini} all of the colours are produced using  just 7~nm thick \gst films. 
The resonant wavelength is changed by controlling the thickness of the ITO layers, hence light is strongly absorbed by the \gst at the resonant wavelength-- in effect, at resonance the  light passes through the \gst film a large number of times, and therefore is strongly absorbed. 
However, at other wavelengths, which are off-resonance, the light is not strongly absorbed. 

\begin{figure}[htbp] 
   \centering
   \includegraphics[width=\textwidth]{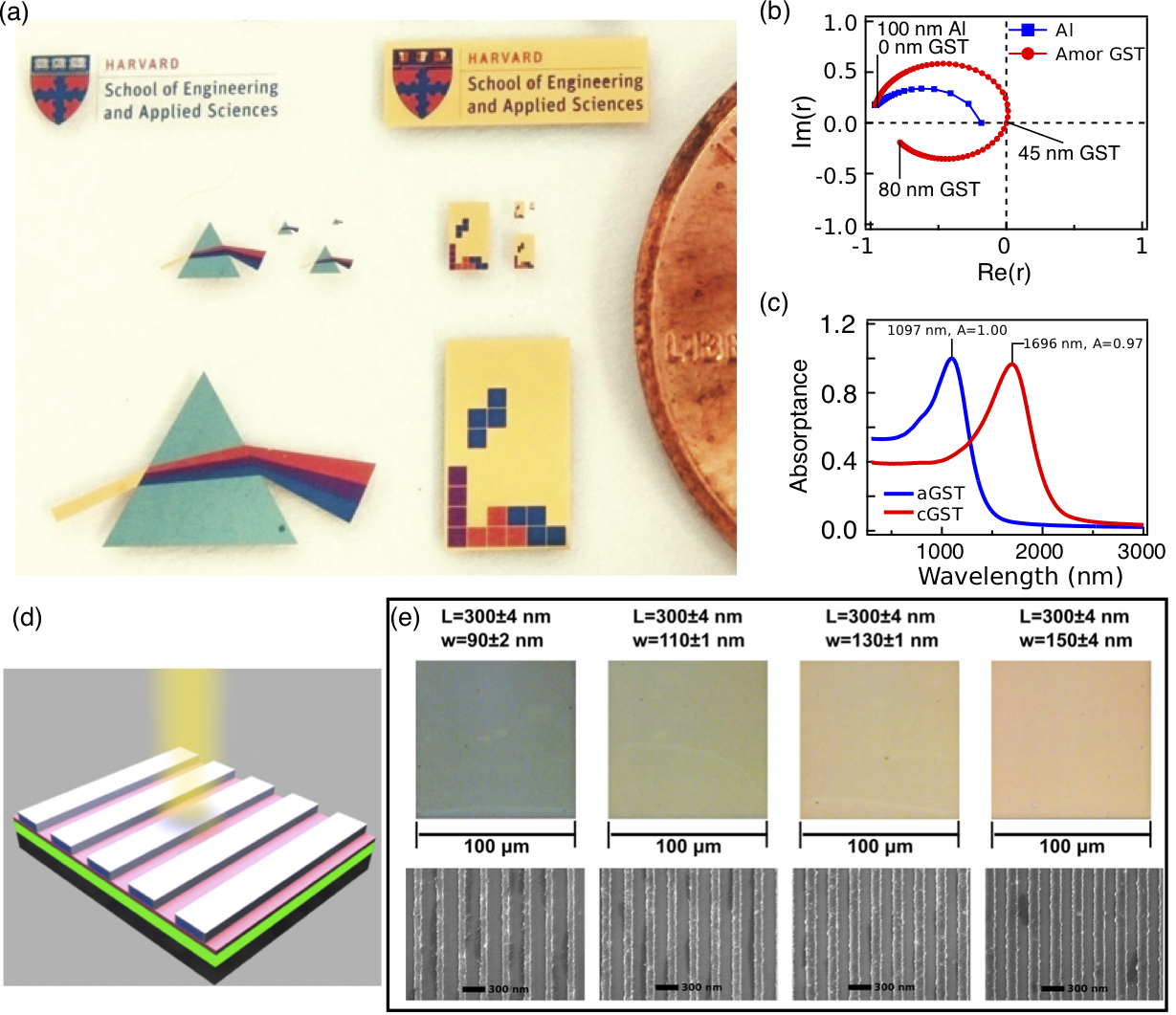} 
   \caption{(a) A photograph showing how ultra thin absorbing Ge films deposited on Au can produce a wide spectrum of reflected colours. 
    7, 11, 15, or 25~nm thick film of Ge were used to produce the 4 different colours.
    See reference \cite{Kats13NMat} for further details.
    (b) The coefficient of reflection is plot as a function of Al (0 to 100~nm) and \gst thickness (0 to 80~nm).
    (c) The modelled reflectivity spectrum for a 45 nm thick \gst film deposited on a 100~nm Al layer.   
    (Please see the supplementary section of ref \cite{Dong18AFM} for further details).
    The strong interference effects cause  resonances at wavelengths of 1097~nm and 1696~nm for the crystalline and amorphous states respectively.
    (d) A schematic of a Al plasmonic grating on top of a \gst \sn layered structure.
    (e) SEM images and corresponding photographs of \gst/\sn/(Al nanograting) structures.
    Figure adapted from references \cite{Kats13NMat} and \cite{Dong16},  Nature Publishing Group and ACS Publications respectively.}
   \label{kats}
\end{figure}

It is also possible to produce strong  resonances by exploiting non-$\pi$~rad phase shifts using extremely thin absorbing films on reflective substrates.
Figure \ref{kats}(a) shows an example of this effect for 7, 11, 15, or 25~nm thick Ge films on a gold  \cite{Kats13NMat}.
Complete destructive interference in these structures is possible by balancing  the film's absorption, with the the amplitude and phase of the the light reflected from each interface.
Replacing the Ge layer with \gst allows the peak absorption wavelength to be tuned over a wide range in the near infrared (N-IR).
However, it is not possible to produce these resonances in the visible spectrum using planar structures.
This is because the \gst absorption coefficient in both crystalline and amorphous films is too large, and therefore the amplitude of  light  after passesing through the thin film structure is insufficient to cancel the light reflected from the Air-\gst interface.
However, at longer wavelength, such as 1097~nm and 1696~nm for the amorphous and crystalline state respectively, the \gst absorption coefficient is lower, and the light reflected from buried interfaces is able to cancel the light reflected from the first air-\gst interface.
This is clearly seen in Fig. \ref{kats}(b-c).
Fig. \ref{kats}(b) shows the complex reflection coefficient as a function of the Al-\gst structure thickness. 
As the Al thickness is increased to 100~nm, the amount of light reflected increases.
Since Al does not strongly transmit light, the path tracks relatively close to the real axis.
At 100~nm Al, the \gst layer is added, which causes the path to change track and follow a clock-wise direction.
The \gst is able to transmit the 1097~nm light, and we start to see the formation of a circular path in the complex plane. 
If the film was perfectly absorbing, then as the film thickness is increased, the reflection coefficient would cyclically track a circular path at a constant radius in the complex plane.
However, \gst has a non-zero absorption coefficient at a wavelength of  1097~nm, hence the the reflection coefficient follows a spiral path.
We see that when the \gst film is 45~nm thick, the path passes through the origin of the complex plane, which means that no light is reflected back from the structure. 
Indeed, this is confirmed in the  Al --\gst(45~nm) structure's absorption spectrum, see Fig. \ref{kats}(c). 
At 1097~nm we see that no light is reflected from the structure.
However, when the \gst is crystallised, the absorption at 1097~nm radically drops from 100\% to aproximately 40\%.
This is because the crystalline \gst film is now more absorbing and the radius of the reflection coefficient  path is now much smaller, hence for the 45 nm thick film, the path does not pass through the origin.
However, at a  wavelength of 1696~nm, crystalline \gst is far less absorbing, and the amount of light absorbed and reflected from the interfaces can balance again to produce near perfect absorption.

It should now be clear from Fig \ref{kats}(b) that  perfect absorption in the visible spectrum is not possible   using dual layer metal--\gst planar structures. 
This is because the absorption of \gst is far too high at visible wavelengths and consequently the reflectance coefficient circle has a radius that is too short to pass through the origin of the complex plane.
This conclusion can also be drawn by examining published reflectance spectra for related structures\cite{Schlich15ACS, bakan2016apl}.
However, other resonating structures, such as plasmonic structures, can be used to enhance the absorption at visible frequencies\cite{Dong16, cao2014SR}.
Fig. \ref{kats}(d) shows one such structure that we implemented. 
Al plasmonic gratings were used to generate plasmon resonances in the visible spectrum.
These resonances couple to the absorbing \gst film and produce strong broadband absorption peaks.
The reflected colour is easily adjusted by controlling the dimensions of the gratings, as seen in Fig. \ref{kats}(e).

The aforementioned  reflective structures  are relatively insensitive to the angle of incidence of the incoming light and can produce large changes in resonant wavelength.
This  is a clear advantage over thicker dielectric stacks that exploit multi-pass light circulation in resonant optical cavities.
The simple planar structure means that they can be manufactured on an industrial scale using well established thin film deposition techniques, such as sputtering.
The main disadvantage is that the peak absorption wavelength is strongly dependent on the intrinsic absorption of the film. 
As was seen, the absorption of \gst is too high to allow completely destructive interference in the visible spectrum.
Clearly, new PCMs need to be developed with low intrinsic absorption at visible frequencies. 
This topic is discussed in section \ref{sec_materials}.

Metamaterials consist of different materials with sub-wavelength features that are patterned to strongly interact with light.
The light `feels' an effective medium that results from the properties of the individual nano materials and their geometric arrangement.
If the metamaterial nanoelements have a resonant response, then the phase of the light emerging from the nanoelement can also be controlled. 
And this allows concepts, such as optical phase arrays to be realised\cite{Yu2014NM}.
To date, most  PCM-tuned metamaterial devices  employ  nanoscale plasmonic  elements.
Since the plasmon resonance of these structures  depends on the refractive index of the surrounding material, the metamaterials' optical response can be tuned by changing the PCM refractive index.
The earliest example of this concept was presented by the Zheludev group.
They used the crystallisation of amorphous Gallium Lanthanum Sulphide\cite{simpson2007electrical} to control  plasmonic split-ring resonators\cite{Samson10apl}.
The same group  also designed a series of \gst-based reconfigurable flat  metasurface devices that refract and diffract light.
A train of femtosecond laser pulses was used to write, erase and over-write  binary and greyscale patterns into the \gst phase change film.
These pronounced two-dimensional  refractive index patterns mould the wavefront of light emerging from the film thus focusing or steering the probe laser beam.
In the following section we will discuss our work on plasmonic metasurfaces, which employ a different strategy to create tuneable infrared filters.

\subsection{Metal--Dielectric--Metal Metasurface}

Metasurfaces are capable of manipulating the phase and amplitude of light without bulky optics.
Hence, they are attractive for compact and highly integrated optical systems. 
The  idea of a metasurface is that rather than being confined to the conventional homageneuos materials, multiple materials can be patterned  with features orders of magnitude smaller than the wavelength of light. 
These nanoscale features are key to the utilisation of phase change materials in  tuneable metasurfaces.
Reversibly switchable \gst-based structures must be quenched at approximately $10^9$ Ks$^{-1}$ for amorphisation\cite{Salinga2013ncom}.
In optical data storage technologies the quenching problem is circumvented  using very thin PCM films.
Indeed, the thermal design of the optical data storage disc structure is extremely important.
A typical disc stack is composed of a polycarbonate substrate/Al (100 nm)/\zsso (20~nm)/\gst (20~nm)/\zsso (140~nm). 
The  \gst layer absorbs the 650~nm laser light and heats in tens of nanoseconds, whilst the other layers remain cool and can quench this thin \gst layer, which has a small thermal load, into an amorphous state. 
Thicker \gst films are very difficult to amorphise because the thermal load is larger.
Therefore, reversibly switchable phase change optical structures must also employ thin PCM layers.
For split-ring resonator metasurfaces, a design compromise needs to be reached.
 This is because reducing the PCM volume produces a smaller change in the metasurface optical response due to the reduced amount of  PCM interacting with the plasmon's electric field.
However, in different metasurface designs, such as metal-dielectric-metal (MDM) metasurfaces,  the  electric field is radically enhanced using gap plasmons in very thin dielectric layers\cite{zhang2005prl, ding2018review}. 
Therefore, we  designed a range of MDM-metasurfaces with GST dielectric layers\cite{cao2018aom,Dong18AOM,cao2014OME,cao2014SR,cao2014SR2,Cao13JOSAB_2,Cao13JOSAB_3,Cao13JOSAB_1,cao2013OME}. 
In these designs the electric field is enhanced in a PCM layer that is just a few tens of nanometers thick, which means the PCM can be quenched at the rates necessary for amorphisation  by using the substrate as a heatsink.
Moreover, the metasurface optical response is  extremely sensitive to changes to the \gst tuneable dielectric constant. 
Since GeTe-\sbte materials strongly absorb light at visible frequencies, we designed our structures to operate at  photon energies below the band gap of the GeTe-\sbte alloys, i.e. in the mid-infrared where the absorption is very low\cite{chew17SPIE}.

\begin{figure}[htbp] 
   \centering
   \includegraphics[width=3in]{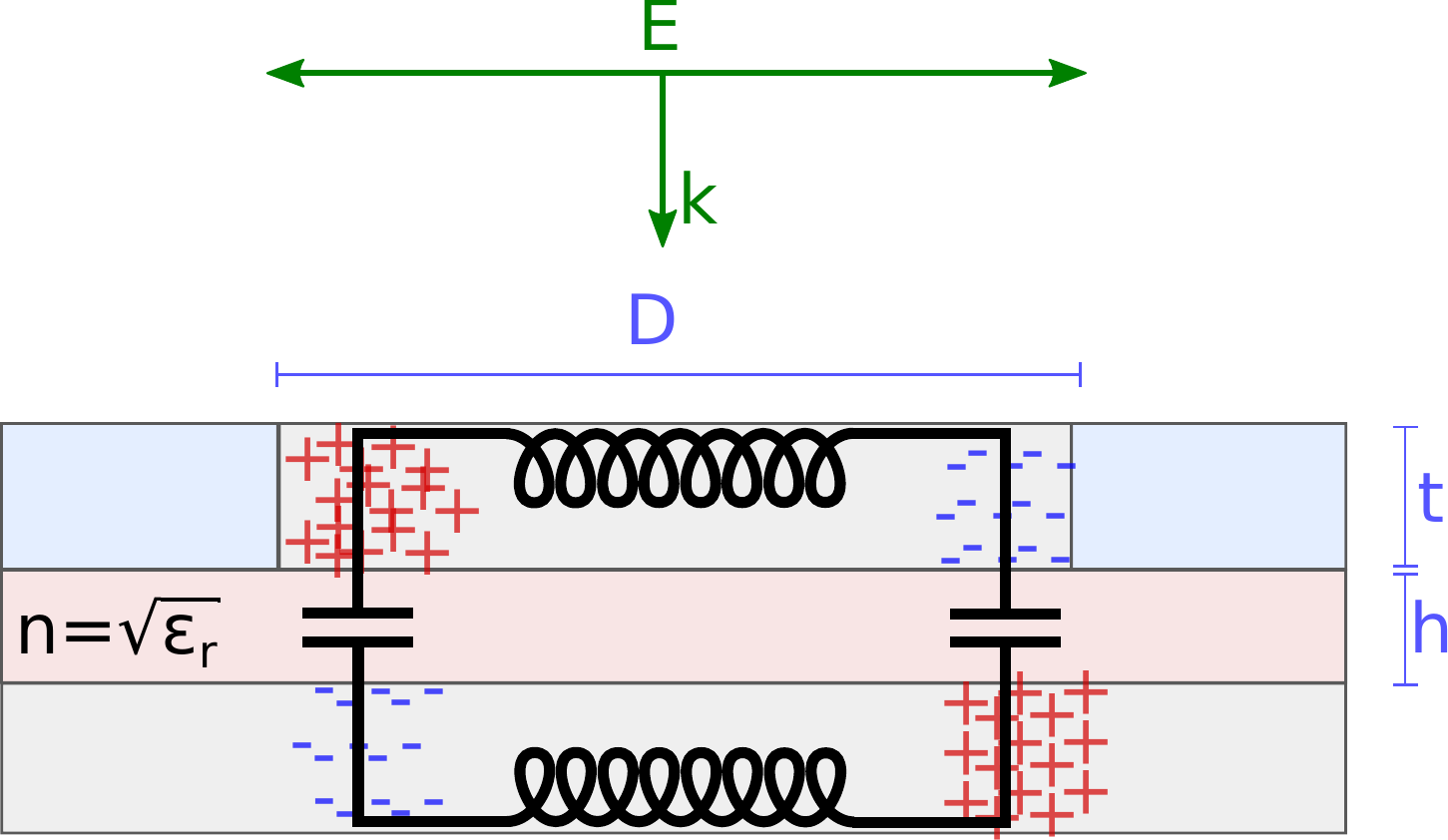} 
   \caption{A metal-dielectric-metal metasurface can be modeled as an LC circuit. The schematic shows how the individual elements of the metamaterial correspond to circuit elements. 
   The grey areas are metallic, the pink area is the dielectric layer, and the blue areas are vacuum.}
   \label{MDMcircuit}
\end{figure}

An electrical circuit model is useful for understanding the behaviour of plasmonic surfaces\cite{Zhu14OE, zhang2005prl, Yang18webinar}.
We can consider the principle parts of the MDM structure as inductor and capacitor elements.
The electric field of the incident light  forces the electrons to move, thus producing an electric current.
This electric current produces a magnetic field; thus, the metallic elements  behave an inductors.
The electrons oscillate due to the driving oscillation of light's electric field.
When the electrons are at their extreme position in the metallic element, they are momentarily static, and there is no magnetic field.
However, now the electrons create image charges in the bottom metallic layer of the MDM structure.
This produces an  enormous electrostatic field in the dielectric layer, and the charge separation can be modelled as a capacitor.
Figure \ref{MDMcircuit} shows a schematic of the MDM equivalent electrical circuit. 
Using this model, we can now describe the nature of the MDM optical response.

When we  design tuneable MDM devices, our main aim is to use PCMs to control the structure's resonant frequency.
Using the circuit model shown in Fig. \ref{MDMcircuit}, we see that the structure can be modelled as an LRC circuit.
The resonant frequency, $f$, of an LC circuit is $f\propto\frac{1}{\sqrt{LC}}$ and therefore we should expect the resonant wavelength $\lambda_{res}$, to be $\lambda\propto{LC}$.
The inductance of a metal slab is correlated to its length and thickness, $L\propto\frac{D}{t}$.
Whilst the capacitance of the MDM structure depends on the dielectric's thickness and permittivity, by $C\propto\frac{\varepsilon D}{h}$.
Therefore, we can now see how the MDM resonant wavelength depends on its geometry and materials, as shown by equation \ref{mdmres}\cite{Yang18webinar}:
 Here, $t$ is the metal layer thickness, $h$ is the dielectric thickness, $n$ is the dielectric refractive index, and $D$ is the length of the inductor.
 
 \begin{align}
 \nonumber
 \lambda_{res}&\propto\frac{\sqrt{\varepsilon} D}{\sqrt{th}} \\
 &\propto n \frac{D}{\sqrt{th}}
 \label{mdmres}
\end{align}

From \ref{mdmres} we  clearly see that the resonant wavelength red-shifts for increasing metal and dielectric film thicknesses and, importantly, when the dielectric's refractive index increases. 
As one might expect, the resonant wavelength is also proportional to the size of the top metal features.

\begin{figure}[htbp] 
   \centering
   \includegraphics[width=\textwidth]{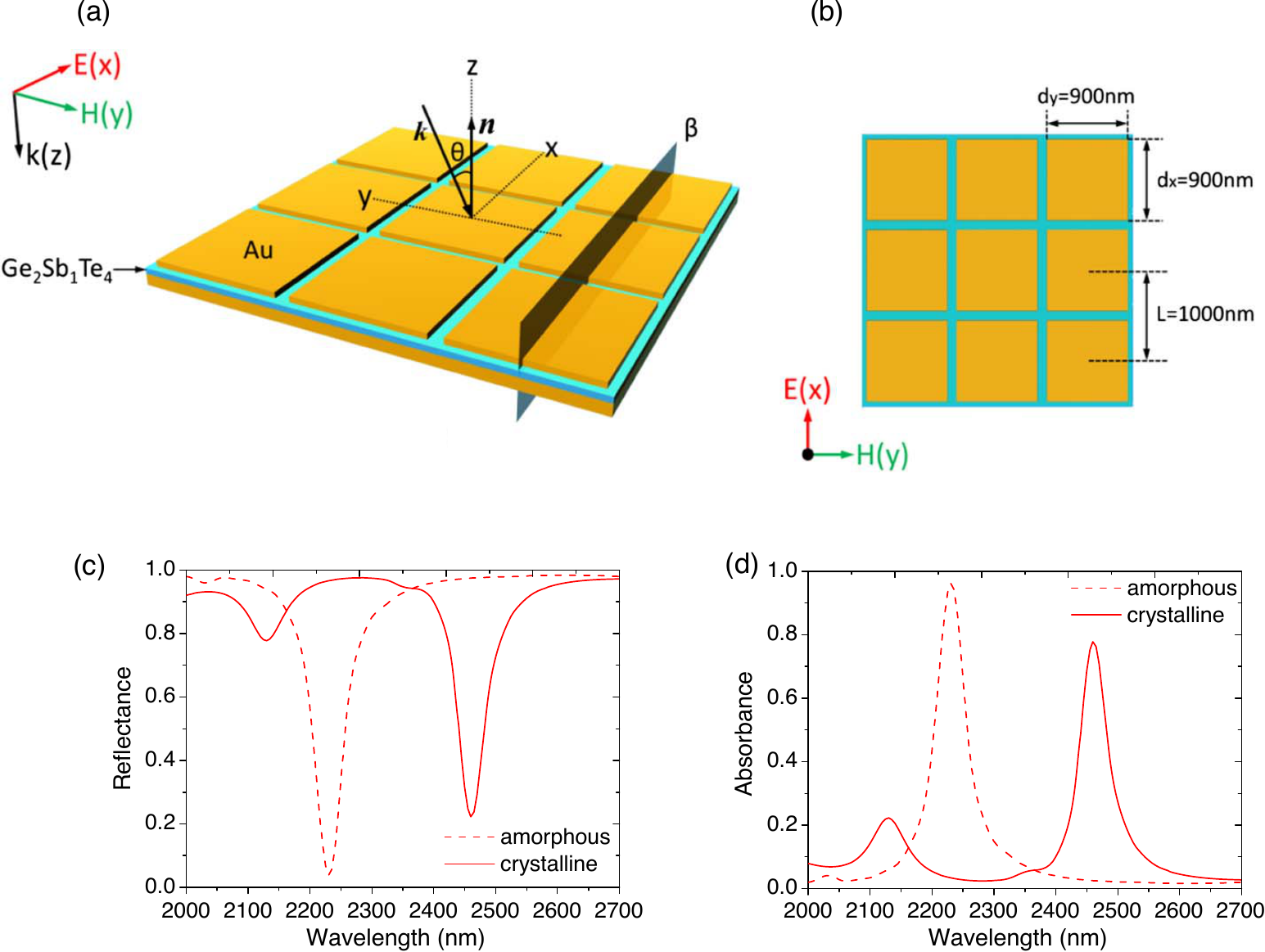} 
   \caption{(a-b) Schematic of the Au-Ge$_2$Sb$_1$Te$_4$-Au MDM metamaterial perfect absorber showing the polarization of the incident light. 
The thicknesses of the Au squares, Ge$_2$Sb$_1$Te$_4$ spacer, and  the Au mirror are 40~mn, 40~nm, and 80~nm respectively. 
The square lattice constant is 1000~nm and the width of Au squares is 900 nm. 
The Finite Difference Time Domain (FDTD) simulations of the reflectance  and absorbance at normal incidence are shown in (c) \& (d) for both the amorphous and crystallines  states of the Ge$_2$Sb$_1$Te$_4$ spacer.
The whole structure was simulated in air. 
Figure adapted from \cite{Cao13JOSAB_3}, OSA Publishing.}
   \label{MDMsqaure}
\end{figure}

If the dielectric layer is replaced with a PCM, such as \gstns, we see from equation \ref{mdmres} that the PCM's large change in refractive index  can be used to tune the  the resonant wavelength of the structure. 
This resonant wavelength shift can be extremely large. 
In 2013  we designed  a Au-Ge$_2$Sb$_1$Te$_4$-Au MDM metamaterial perfect absorber that demonstrates this wide tuneablility\cite{Cao13JOSAB_3}. 
The schematic of our design is shown in Figure \ref{MDMsqaure}(a-b). 
Essentially it consists of square array of 40~nm thick Au squares, which are pattern atop of a 40~nm thick continuous Ge$_2$Sb$_1$Te$_4$ film, which in turn is on top of a 80~nm thick continuous Au film. 
The square lattice has a 1000~nm lattice constant, with Au squares with 900~nm long sides.

At resonance the absorption of the MDM Au square array metamaterial increases dramatically.
This is because the effect of a small amount of absorption is amplified as the energy cycles back and forth.
Thus, at resonance,  troughs in the reflection spectrum are observed (see Fig.  \ref{MDMsqaure}(c)), whilst peaks are seen in the absorption spectrum (see Fig. \ref{MDMsqaure}(d)).
When the Ge$_2$Sb$_1$Te$_4$ layer is in the amorphous state, the strongest resonance is at $\lambda_{res}=2230$~nm.
However, if the Ge$_2$Sb$_1$Te$_4$ layer is crystallised, the resonance shifts to $\lambda_{res}=2460$~nm due to the increased refractive index.
This switching behaviour is  predicted by equation \ref{mdmres}.

\begin{figure}[htbp] 
   \centering
   \includegraphics[width=\textwidth]{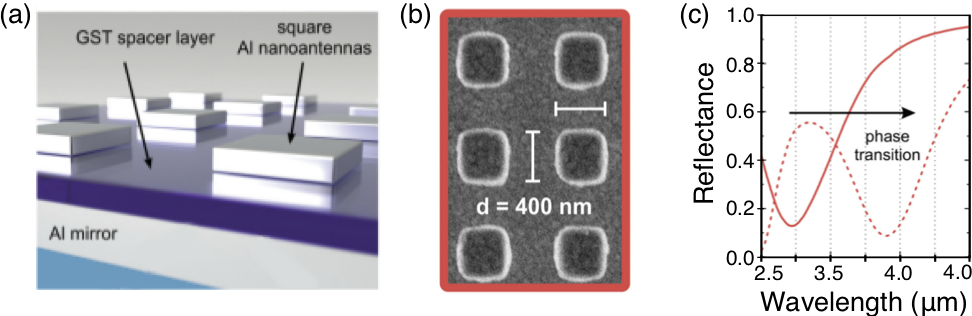} 
   \caption{(a) Schematic of the prototype perfect absorber with phase change material tuned resonances in the mid-infrared. 
   The square array of antennae are  stacked above a Ge$_3$Sb$_2$Te$_6$ spacer and an Al mirror. 
   (b) Scanning electron microscope image of the square aluminium nanoantennas. 
   The Al squares have a length of 450~nm.
   (c) The corresponding measured reflectance spectra for the amorphous and crystalline states of the Ge$_3$Sb$_2$Te$_6$ spacer. 
Figure adapted from \cite{tittl15AM}, OSA Publishing. }
      \label{tittl}
\end{figure}

This type of MDM mid-infrared tuneable perfect absorber  was successfully verified in 2015\cite{tittl15AM}.  
An array of square aluminium (Al) resonators stacked above a Ge$_3$Sb$_2$Te$_6$ spacer layer and a bottom Al mirror was used to almost perfectly absorb mid-infrared light.
Impressively, the absorption resonances could be red-shifted by 700~nm from 2700~nm to 3400~nm by crystallising the Ge$_3$Sb$_2$Te$_6$ at 180\textdegree{}~C.
This is shown in fig. \ref{tittl}

For GeTe-\sbte--based photonic resonators the Q-factor of the resonance decreases when the PCM crystallises.
This is because when GeTe-\sbtens  PCMs crystallise, their electrical conductivity radically increases.
Considering the LC circuit model (Fig. \ref{MDMcircuit}) once again, we  see that if the  PCM becomes conductive,  a `shunt' resistor forms in parallel with the capacitors.
This, in turn, means that charge can easily leak away from the capacitors  through their PCM spacer. 
Thus energy is lost from the resonator, and the Q-factor of the resonator drops.
Indeed, slightly broader resonances are seen for the crystalline state in the measured absorption spectra presented by Tittl et al\cite{tittl15AM}.

\begin{figure}[htbp] 
   \centering
\includegraphics[width=\textwidth]{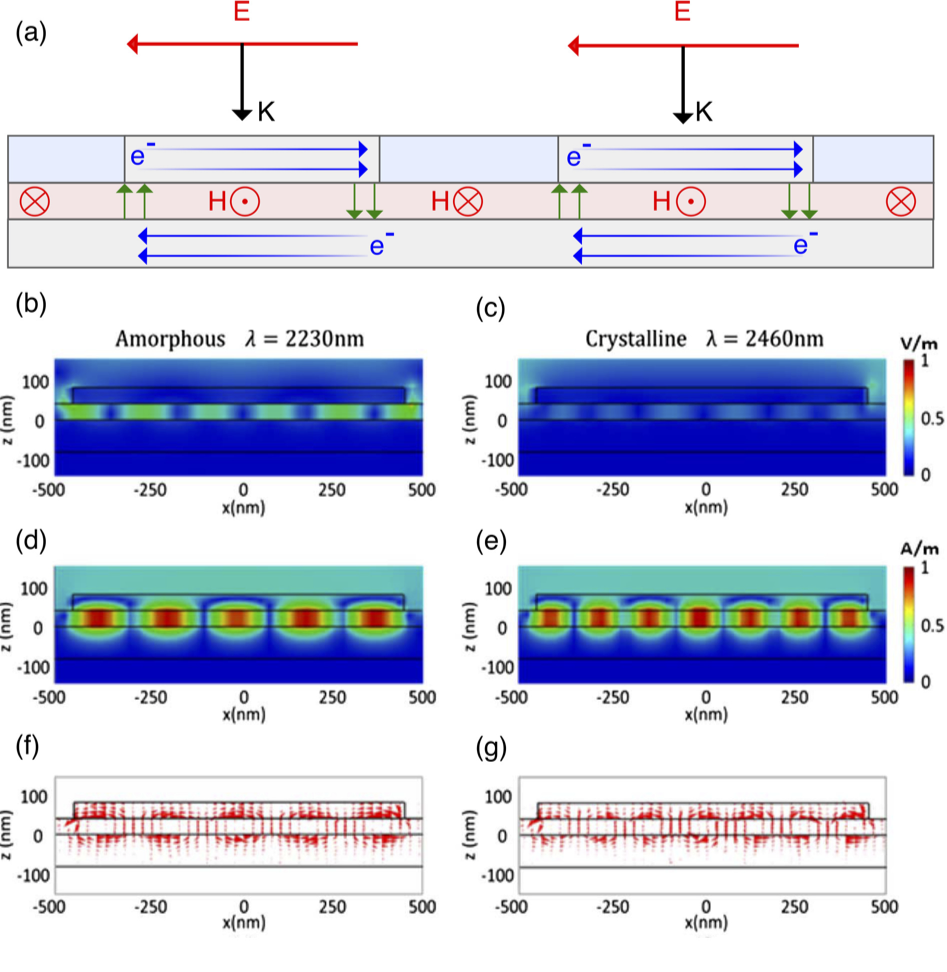}
   \caption{(a) Schematic of the MDM perfect absorber structure at resonance. 
   Current loops (blue and green arrows) are generated by the electrons, which are forced to move in the conductor arrays by the incoming E-field.
   These current loops generate magnetic dipoles that strongly absorb the energy of the incoming magnetic field.
 The FDTD simulated total electric field intensity  (b-c), total magnetic field intensity  (d-e), and  displacement current  distributions are shown for the amorphous  and crystalline Ge$_2$Sb$_1$Te$_4$ MDM perfect absorber structure at resonance, see Fig.\ref{MDMsqaure}.
 The figure was adapted from ref \cite{Cao13JOSAB_3}, OSA Publishing.}
   \label{currentloops}
\end{figure}

In order for these MDM structures to perfectly absorb light,   both the electric and magnetic components of the incoming light must be efficiently absorbed.
The incoming electric field forces free-electrons to move in the conductors, which are patterned on the surface of the MDM.
As the electrons move, they induce image charges in the bottom conducting layer of the MDM structure, which move in the opposite direction.
Fig. \ref{currentloops}(a) shows a schematic of the MDM structure with electric field ($\vec{E}-$), magnetic ($\vec{H}$-) field, and charge movements indicated. 
In the figure, charges moving in the metal layers are indicated by blue arrows, whilst displacement currents are highlighted by green arrows.
The charge  at the edge of the conducting squares is constantly changing with respect to time due to the cyclical nature of the incoming E-field.
Thus a displacement current is generated through the dielectric spacer layer, and a current loop consisting of metallic and displacement currents is formed.
These current loops generate a magnetic dipole, and consequently a magnetic dipole moment is also generated due to the incident H-field. 
Therefore, the H-field also strongly interacts with the structure, and is absorbed at resonance. 
If the structure is designed such that the H-- and E-- fields are resonant at the same frequency, then the incident light is perfectly absorbed.
Figure \ref{currentloops}(b-g) shows this effect for the phase change MDM structure presented in Fig.\ref{MDMsqaure}.
At resonance the $\vec{E}-$ and  $\vec{H}-$ fields are tightly confined in the Ge$_2$Sb$_1$Te$_4$ spacer layer. 
This is easily seen in the E-field (Fig. \ref{currentloops}(b-c)) and H-field intensity (Fig. \ref{currentloops}(d-e)) maps.
The corresponding current loops are shown in Fig. \ref{currentloops}(f-g).
Strong magnetic moment generating current loops are present at resonance, which causes the observed strong peaks in Fig. \ref{MDMsqaure}   

\begin{figure}[htbp] 
   \centering
\includegraphics[width=\textwidth]{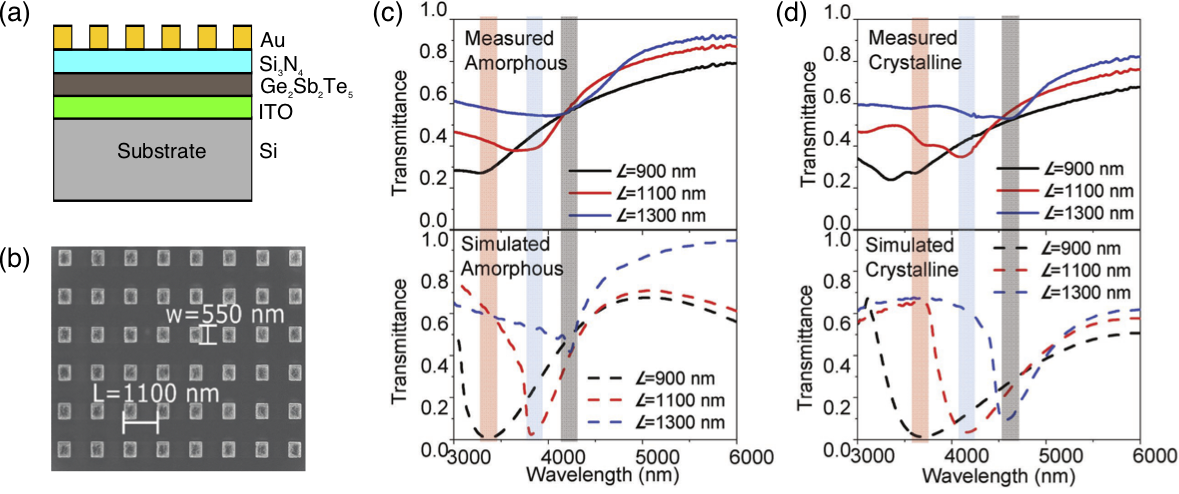}
   \caption{(a) A schematic of the MDM transmissive filter, (b) SEM image of the Au square array surface of the structure. 
   (c) and (d) respective show the transmission spectra for the amorphous and crystalline states of the MDM metasurface. 
   A series of transmission spectra measurements and simulations are given for varying pitch:  L=900~nm (black), 1100~nm (red), and 1300~nm (blue). 
   Each resonator had a square side-length of 550~nm.
Adapted from \cite{Dong18AOM}, Wiley. 
   }
   \label{ITO}
\end{figure}

Most  MDM metasurfaces are designed to reflect light, rather than transmit it.  
However transmission-type filters are attractive because they can be placed directly in front of detectors to create spectroscopy systems.
These in-line designs are simpler  than the reflective types, and consequently they can be used in highly compact integrated systems-- such as in front of a camera CCD. 
However, the MDM structure usually consists of a bottom metallic reflector, which is required to create plasmon-image charges. 
Hence, if conventional materials are used, these devices cannot be transmissive.
To solve this design-conflict we used Indium Tin Oxide (ITO), which is a transparent conductor, as the bottom electrode.
The absorbtion of ITO can be tuned to become somewhat transparent in the M-IR by changing the oxygen content\cite{lee1993jvst}.
As is seen in Fig. \ref{ITO}, the transmission spectra shows resonances in the IR, which can be tuned by changing the phase of the \gstns.
Note, the resonances are  broad because these structures are  lossy due to the relatively high transparency of the bottom ITO layer.

We believe  MDM metasurfaces have a high potential for real practical application because the PCM layer can be very thin, which is necessary for reversible switching, and only  a single lithographic step is required to pattern the devices. 
Although electron beam patterning is impractical for large area industrial scale manufacturing, other \gstns-tuned MDM structures that employ self-assembled  plasmonic nanoparticles also show tuneable near perfect absorption\cite{Cao19ami}, and we believe these can provide a  low cost and high throughput route to commercialisation.

\section{Material Development}
\label{sec_materials}

The objective of this section is to introduce new materials that are being developed  specifically for tuneable phase change photonics applications.
New chalcogenide PCMs and alternative plasmonic materials will be discussed.
In particular we will introduce  \SbS as a PCM for tuneable visible photonics, and discuss issues associated with diffusion and chemical reactions between chalcogenide PCMs and  plasmonic metals.

PCMs, such as  Ag-In-Sb-Te and those along the GeTe--\sbte{} pseudo-binary tie-line, are commercially proven technologies in re-writeable compact discs, digital versatile discs, and blu-ray discs.
This is due to their large  refractive index contrast between structural states\cite{yamada91}, fast switching speed\cite{Waldecker2015nmat},  scalability\cite{Simpson2011nn}, availability, and proven commercial success in data storage applications.
Clearly, it is attractive to start phase change photonics research with a technologically proven PCM, such as \gstns.
However, \gst has a band gap of 0.5~eV in the crystalline state and 0.7~eV in the amorphous state\cite{lyeo2006apl}.
This means that it is highly absorbing in the visible and in the near infrared part of the electromagnetic spectrum.
\gst{} only becomes transparent for wavelengths beyond $2\mu m$ where it retains a substantial change in the real component of the refractive index. 
This is most clear in Fig. \ref{gstnk}, which shows the measured refractive index of \gst{} to a wavelength of 3 $\mu$m and the extrapolated refractive out to a wavelength of 6 $\mu$m. 
The refractive index measurement and analysis details are given in reference \cite{chew17SPIE}, whilst the measurement data is open and available on \href{www.actalab.com}{www.actalab.com}.

\begin{figure}[htbp] 
   \centering
   \includegraphics[width=0.8\textwidth]{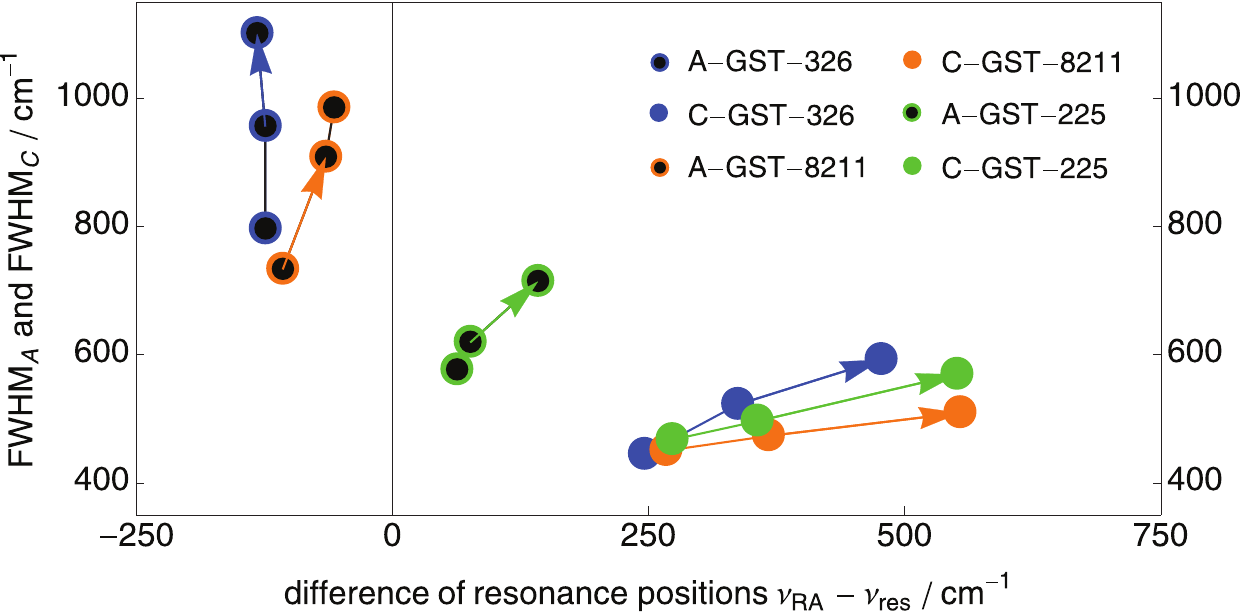} 
   \caption{The resonance full width half maximum (FWHM) for antennae of different lengths (628 $\pm$ 20, 517 $\pm$ 20, and 418 $\pm$ 20 nm) and  different \sbte-GeTe compositions in different structural states are plot against the difference between the antennae resonant frequency shift.
Reproduced from ref \cite{michel2017aom}, Wiley }
   \label{Michel}
\end{figure}

Despite the popularity of \gst for tuneable M-IR photonics, there may be other \sbtens--GeTe compositions that are better suited to a particular application.
Indeed, the Taubner group analysed the infrared optical properties of the \sbtens--GeTe pseudo-binary alloys with the aim of designing noble metal plasmonic resonators that exhibit sharp resonances in the infrared\cite{michel2017aom}.
The Q-factor of any resonator is inversely proportional to the energy lost by the resonator during one cycle.
It is, therefore, important to identify \sbtens--GeTe compositions with a low-loss in the IR.
Fig. \ref{Michel} shows the resonance Full Width Half Maximum (FWHM) for different antennae lengths ($628~\pm~20$, $517~\pm~20$, and $418~\pm~20$ nm), different \sbte--GeTe compositions, and  different structural states of the PCM.
These FWHMs measurements are plot against the difference between the antennae resonant  frequency with, and without the PCM. 
As can be seen,  \gst exhibits the smallest shift in resonance frequency ($\nu_{RA}-\nu_{res}$).
In comparison,  the Ge$_3$Sb$_2$T$_6$ and  Ge$_8$Sb$_2$Te$_{11}$ compositions offer a much broader tuning range; almost twice that of \gstns.
 Ge$_3$Sb$_2$T$_6$ has the largest resonant bonding character of the alloys studied, and therefore it is not surprising that antennae   tuned by it  show a large spectral change\cite{michel2017aom}.
Optical losses in these structures are seen as a widening of the FWHM. 
In almost all the measurements, the Ge$_3$Sb$_2$T$_6$  composition has the highest loss, whilst Ge$_8$Sb$_2$Te$_{11}$ has the lowest loss.
This is because  Ge$_8$Sb$_2$Te$_{11}$ has the largest band gap of the three materials, which sets the short wavelength absorption edge\cite{michel2017aom}.
Its small carrier concentration also lowers the losses at IR frequencies associated with the Drude response\cite{michel2017aom}.
The combination of lower losses and large tuning range means that the  Ge$_8$Sb$_2$Te$_{11}$ alloy has the largest tuning figure of merit, which is defined as, $TFOM=\frac{\Delta \nu}{FWHM}$.
 $\Delta \nu$ is the change in the resonator's frequency between amorphous and crystalline states, and FWHM is the full width half maximum of the resonance peak.
Therefore, of all the PCMs along the \sbte-GeTe pseudobinary tie-line, Ge$_8$Sb$_2$Te$_{11}$  offers the best compromise between  Q-factor and  tuning range. 
It is, therefore, recommended for tuning M-IR devices.

\begin{figure}[htbp] 
   \centering
   \includegraphics[width=.9\textwidth]{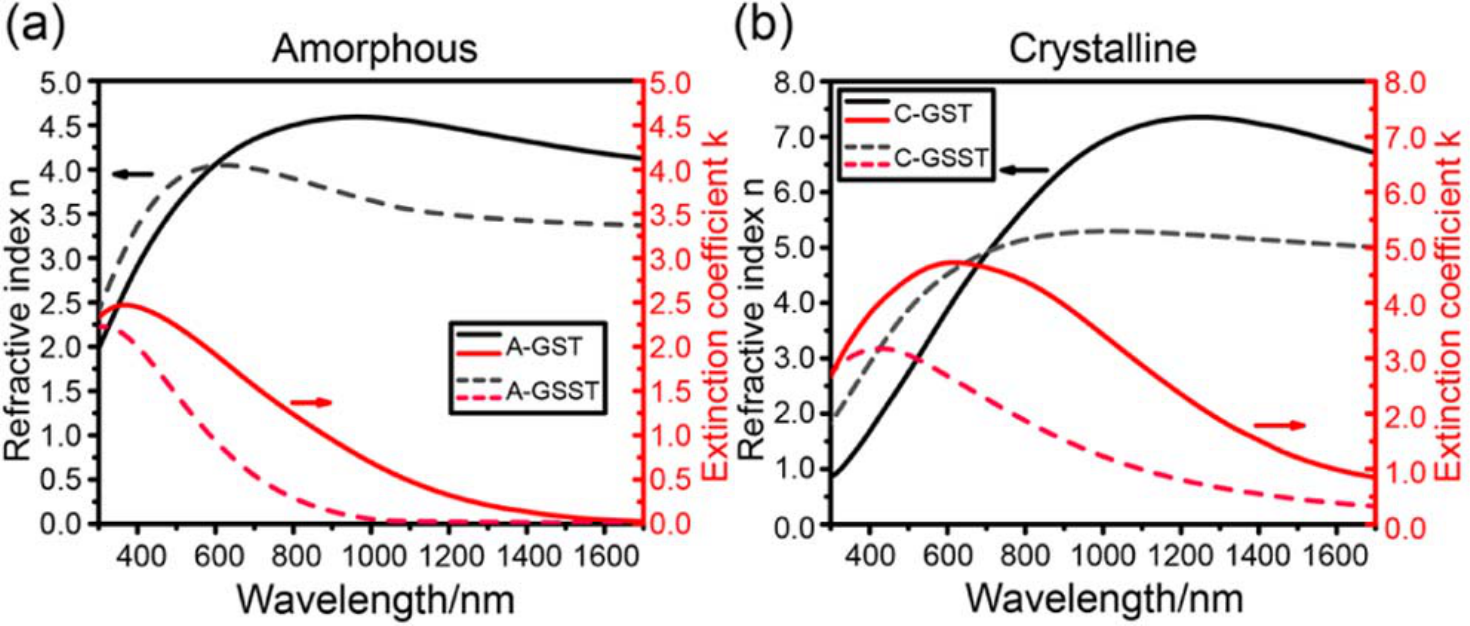} 
   \caption{A comparison of the ellipsometry measured optical constants for Ge$_2$Sb$_2$Se$_4$Te$_1$  and \gst in the amorphous and crystalline states.
 Reproduced from ref \cite{Zhang17OL}, OSA Publsihing.}
   \label{GSST}
\end{figure}

Optical losses for \sbtens--GeTe compositions in the N-IR and visible are extremely high, and therefore new PCMs must be developed for these frequency bands.
The Hue group has developed the Ge$_2$Sb$_2$Se$_4$Te$_1$ by taking the \gst alloy composition and substituting 80\% of the Te with Se\cite{Zhang17OL}. 
The optical constants for this material are compared with \gst in figure \ref{GSST}.
As can be seen, even after substituting the Te with Se, there is still a large change in Re(n), whilst  the extinction coefficient, Im(n), is substantially lower in both crystalline and amorphous states.
Although the optical band gap of the material is not given, we  see  that the amorphous state transmits wavelengths longer than 1~$\mu$m. 
The crystalline state seems to have a long Urbach tail but is significantly less lossy than \gst in the telecoms band.
Early preprint results also suggest that this material can reversibly switch on a microsecond timescale and be cycled more than 1000 times between the crystalline and amorphous states\cite{zhang2018axiv}.

\begin{figure}[htbp]
   \centering
   \includegraphics[width=0.7\textwidth]{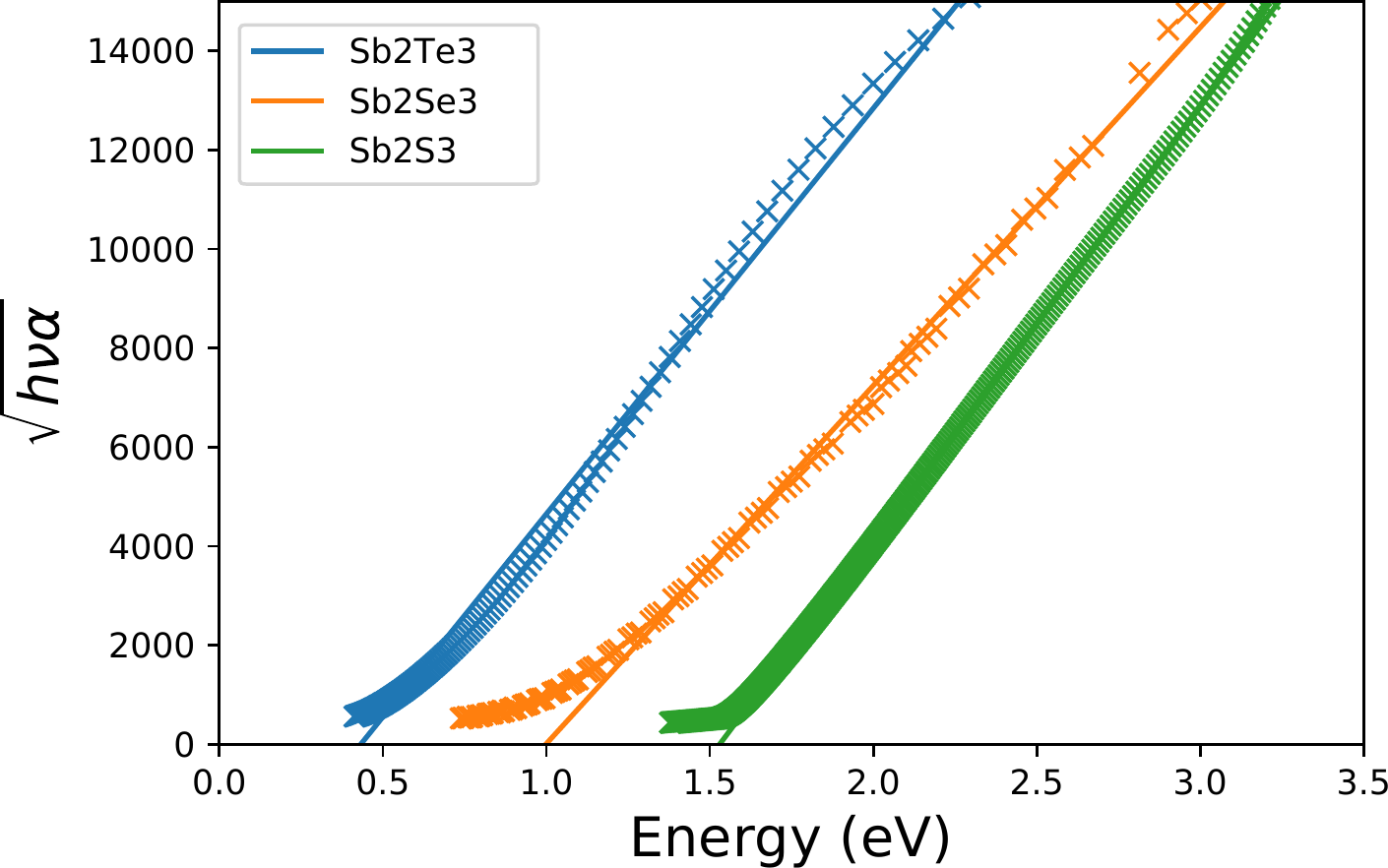}
   \caption{Tauc plots for Sb$_2$S$_3$, Sb$_2$Se$_3$, and \sbte{} crystalline films. The data was taken from references \cite{Chen15apl, Dong18AFM}  }
   \label{Sb2M3}
\end{figure}

It is not surprising that substituting Te with Se in \gst increased its optical band gap. 
As a rule of thumb, as lighter chalcogen atoms are replaced  by heavier chalcogen atoms, moving down the periodic table from S to Se to Te, the band gap of the chalcogenide  becomes smaller.
 Eventually the material becomes a metal or semi-metal. 
This is  clearly seen for the Sb$_2$M$_3$ system, where M is either S, Se, or Te.
We see from the Tauc plot, which is shown in Fig. \ref{Sb2M3}, that the optical band gaps for Sb$_2$S$_3$, Sb$_2$Se$_3$, and \sbte{} are 1.5~eV, 1.0~eV, and 0.43~eV respectively\cite{Chen15apl, Dong18AFM}. 
Since each of these materials  stably exhibit optical contrast between their amorphous and crystalline states, these  materials can be used as the basis to develop new PCMs for specific spectral regions.

\begin{figure}[htbp] 
   \centering
   \includegraphics[width=\textwidth]{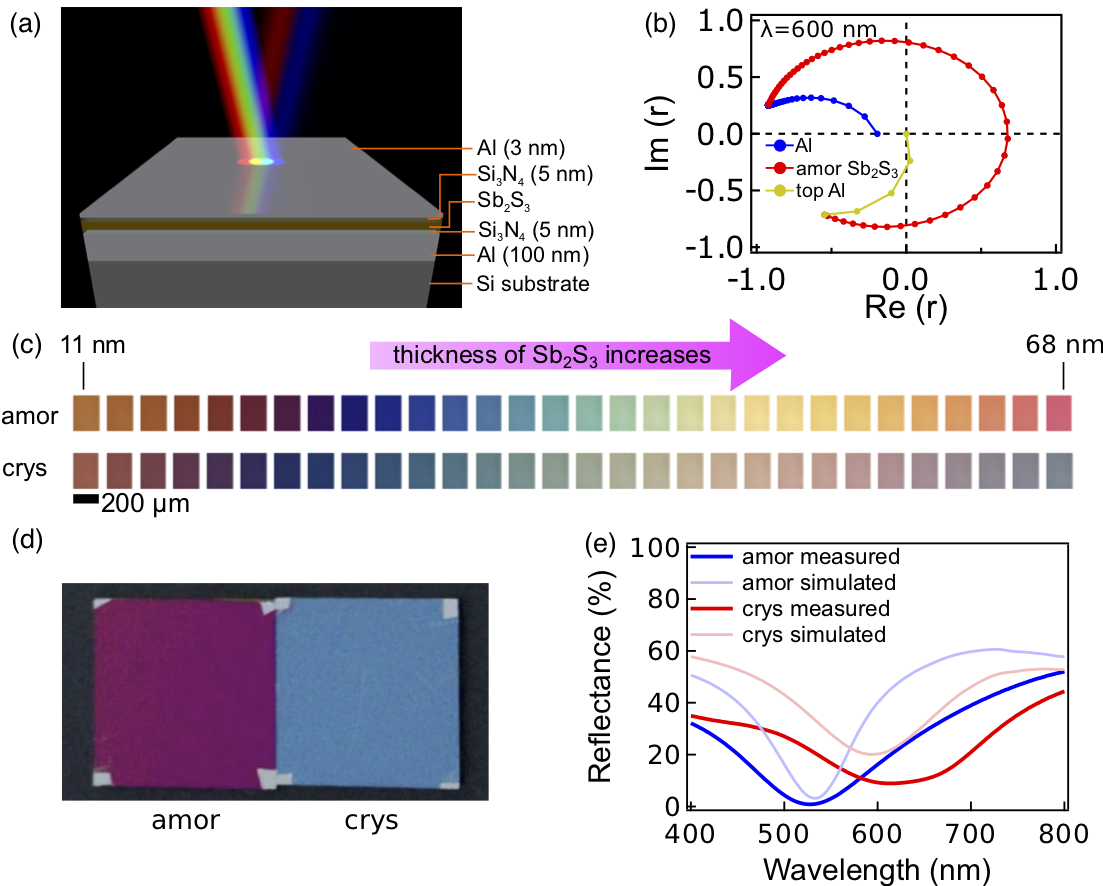} 
   \caption{(a) A schematic of the Al-\SbS tuneable filter. 
   b) The complex reflection coefficient for 600~nm light. 
   c) Photographs of the colours produced for different thicknesses of \SbSns ranging from 11 to 68~nm.  
   (d) Photograph of the structure in the amorphous and crystalline states. 
   e)  Measured and simulated reflectance spectra for the structure  with a 24~nm thick \SbS layer in the amorphous and crystalline states.
 Figure adapted from ref \cite{Dong18AFM}, Wiley.  }
   \label{DongAFM}
\end{figure}

We aimed  to use phase transition induced changes to the absorption coefficient to radically tune the colour of metal-\SbS filters in the visible spectrum.
We chose \SbS  because of  its large band gap and concomitant lower absorption coefficient in the visible spectrum.
We also found that this material is capable of  reversible switching. 
It crystallises from the as-deposited  amorphous  state in just 70 ns\cite{Dong18AFM}.
This means that material can be used for high speed reprogrammable photonics applications.
In section \ref{planar} we saw that when extremely thin absorbing semiconductor films are deposited on reflective metal films, perfect light absorption is possible via strong interference effects.
However, the perfect absorption wavelength depends on the absorption coefficient of the film, and the intrinsic optical absorption of \gst in the visible spectrum is too large to allow perfectly destructive interference. 
However, \SbS has a much lower absorption coefficient at visible frequencies.
Moreover, the absorption edge shifts from 2.05 to 1.70~eV  when the material crystallises\cite{Dong18AFM}.
We designed, therefore, a metal--\SbS thin film stack structure with a resonant frequency that  couples at photon energies in the 2.05 to 1.70 eV range. 
Fig. \ref{DongAFM}(a) shows a schematic of the structure.
We also included \sn barriers to prevent any reactions between \SbS and the Aluminium reflector layers. 
A 4~nm thick Aluminium top layer was also used to enhance the destructive interference.
Indeed, as can be seen in Fig. \ref{DongAFM}(b), the top Aluminium layer causes the reflectance coefficient to pass through the origin of the complex plane circle diagram; thus allowing perfect destructive interference. 
Without this layer, the reflectivity at resonance will be much higher.
Fig. \ref{DongAFM}(c) shows that changing the thickness of the \SbS layer can produce a wide variety of vibrant colours. 
The \SbS phase transition produces the largest spectral shift when the film is 24~nm thick.
The colour changes from pink in the as-deposited amorphous phase to blue in the crystalline phase for the 24~nm thick \SbS film. 
Photos of the film are included in Fig. \ref{DongAFM}(d), whilst the corresponding reflectance spectra are shown in Fig. \ref{DongAFM}(e).
Note, that the vibrancy of the colours produced by combing metal and \SbS films is due to  low losses in the visible spectrum. 
When losses are high, the resonances are broader, which spreads the reflected wavelength across many colours thus causing a `greying' effect.

\begin{figure}[htbp] 
   \centering
   \includegraphics[width=0.8\textwidth]{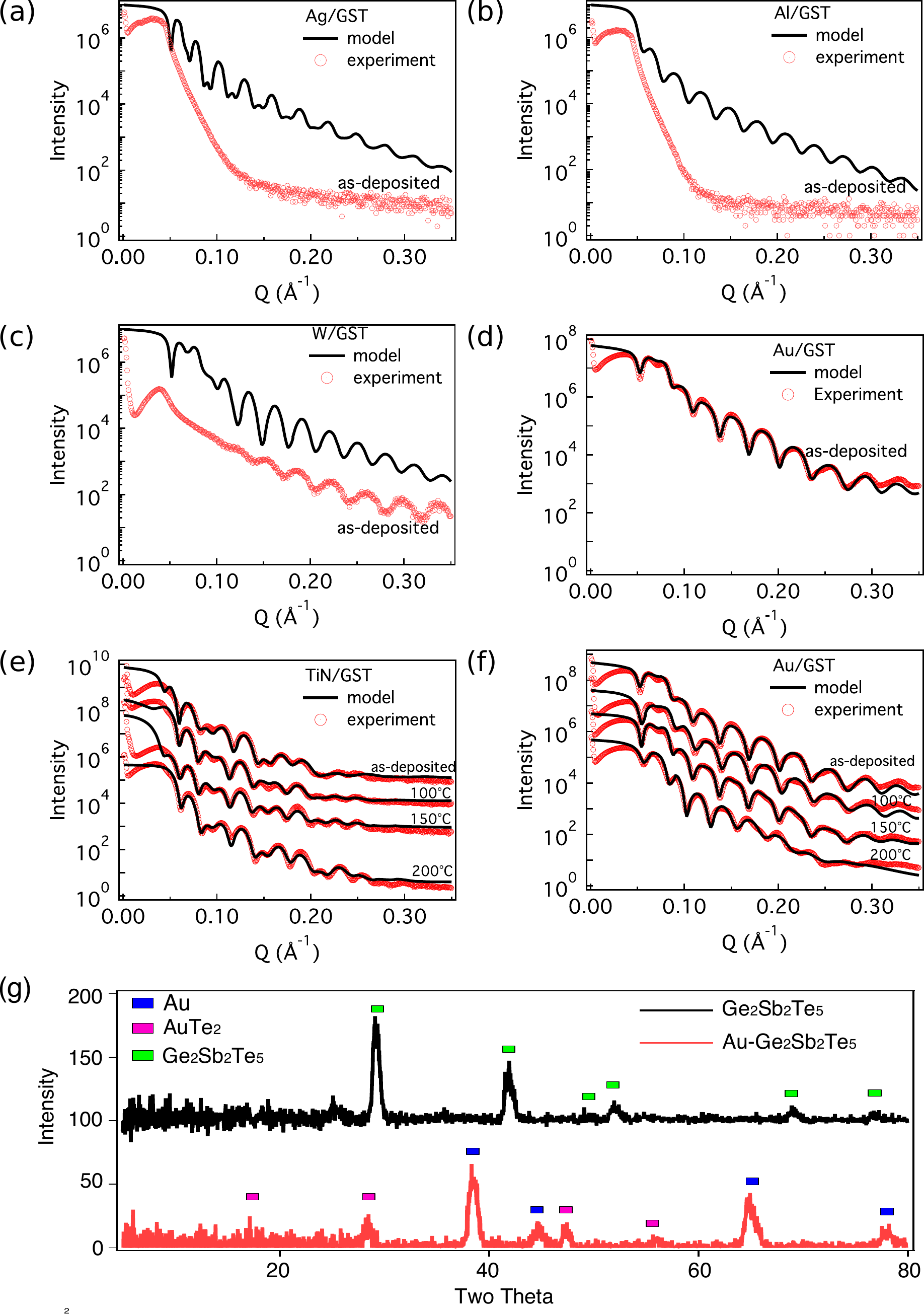} 
   \caption{X-Ray Reflectivity and X-Ray diffraction measurements and models exemplafying the problems with interdifssion and reactions at \gst-metal interfaces.
    (a-d) Metal--amorphous \gst interfaces, where the metal is (a) Ag, (b) Al, (c) W, and (d) Au. 
    (e-f) show  TiN/\gst and Au/\gst interfaces after annealing at 100  \celsiusns,150  \celsiusns, and 200  \celsiusns.
    Figure adapted from ref \cite{lu2018JMS}, Springer Publishing}
   \label{XRR}
\end{figure}

Aside from PCMs, it is also important to consider the materials that interface with the PCM. 
We have found that many plasmonic metals react or diffuse into \gstns \cite{lu2018JMS}.
This is clearly seen in Fig. \ref{XRR}(a-f), which shows the measured X-Ray Reflectivity (XRR) plots of \gst interface with different metals after annealing at different temperatures.
The figure also includes the modelled XRR plots assuming that the interfaces are perfectly sharp.
Silver, Aluminium, and Tungsten all seem to react with, or diffuse into, \gst during the sample growth.
From the XRR study, gold appears to be stable up to 200~$^\circ$C but the X-Ray diffraction measurements (see \ref{XRR}(g)) clearly show the formation  of AuTe$_2$.
We found that this reaction  destroys resonances in thin film Au-\gst filters\cite{lu2018JMS}.
To prevent these reactions, a diffusion barrier, such as a 5~nm thick \sn film between the \gst and gold, can be used.
However, adding  barriers between the PCM and plasmonic films is undesirable because it decreases the optical change induced  by the PCM's phase transition.
Indeed, the PCM should be placed in direct contact with the plasmonic metal to maximise its overlap with the tightly confined surface plasmon electric field.
Thus, we recommend  TiN  as the plasmonic material for PCM tuned plasmonic applications. 
We see that the TiN--\gst XRR pattern shown in Fig. \ref{XRR}(e) almost exactly fits an XRR model of TiN/\gst with sharp and smooth interfaces. 
This suggests that the measured layers have neither reacted nor inter-diffused.
Importantly, the TiN   visible dielectric function is very similar to that of gold\cite{Naik12OMEx}, and TiN is a proven electrode heater material  in Phase Change Random Access Memory devices\cite{burr2010jvstb}.
The fact that the \gst in PCRAM devices can be heated above the \gst melting temperature millions of times when the devices are switched suggests that  that  TiN-\gst interfaces are extremely stable.

\section{Summary and Outlook}

A variety PCM-tuned filters, metasurfaces, and On-Chip devices have been designed and prototyped.
Often these devices exploit the well-known data storage PCMs along the \sbte-GeTe pseuduo-binary tie-line.
We believe these particular compositions are most useful for tuning M-IR devices, where optical losses are low.
Most of the current M-IR research focusses on the \gst composition, however, detailed experimental studies have shown that Ge$_8$Sb$_2$Te$_{11}$ exhibits the best compromise between optical losses and tuning range\cite{michel2017aom}.

Since these PCMs require high quench rates to switch into the amorphous states, most of the prototype device demonstration only show one-way switching from the amorphous state to the crystalline state. 
A significant challenge is  designing PCM-based photonic devices that can be reliably cycle between their amorphous and crystalline states.
MDM structures use very thin PCM films to produce enormous changes to their optical response, whilst the PCM's small thermal load also permits short quench times.
For this reason, and because of their relatively simple design,  we believe that MDM structures show the greatest promise for commercial application.

Aside from \sbte-GeTe PCMs, a number of new PCM compositions are being developed specifically for photonics applications.
These materials tend to replace Te with S or Se. 
In particular we have focussed on \SbSns, whilst others are exploring Ge$_2$Sb$_2$Se$_4$Te$_1$.
Both materials have a significantly larger optical band gap than the \sbte-GeTe alloys, and they can be reversibly switched using sub-microsecond Joule heat pulses\cite{Dong18AFM, zhang2018axiv}.

When designing PCM-tuned photonic devices, it is also important to consider chemical reactions and inter-diffusion with the interfacing materials.
For example, we have found that many plasmonic materials  diffuse into or react with \gstns \cite{lu2018JMS}.
Thus when designing tuneable plasmonic devices in the visible spectrum, we recommend using TiN for the plasmonic material.
It can be interfaced directly with the PCM without reactions of inter-diffusion, and it can be used as the heating electrode for switching the material.
If, other plasmonic materials must be used, then a thin barrier of an inert material should be included between the plasmonic metal and the PCM.

In conclusion, many PCMs exhibit tuneable electrical and  optical properties, which have been exploited in electrical and optical memory technologies.
However, we believe the next generation of PCM-based devices will exploit the electrical and optical properties simultaneously.
There will be a growing need for these types of materials as their application is not limited to a single specific task; i.e. they are reconfigurable and may  perform  variety of photonic and electrical tasks.
Thus, PCMs  have the necessary plasticity to fuel smart multifunctional devices, systems, and networks.
These types of devices will be an important-enabler of the Internet of Everything

\emph{Acknowledgements. We thank Professor Joel K. Yang for enjoyable and fruitful discussions. 
We are grateful for the hard work of our PhD students: Weiling Dong, Li Lu, Mr Libang Mao, and Kuan Liu. 
A substantial portion of the results presented in this chapter are the outcome of  A-Star (A18A7b0028, 1420200046) and the Samsung Global Research Opportunity grants. }

\end{document}